# Emulating Batteries with Deferrable Energy Demand
## Fundamental Trade-offs and Scheduling Policies

Daria Madjidian, Mardavij Roozbehani, and Munther A. Dahleh

*Abstract*—We investigate the ability of a homogeneous collection of deferrable energy loads to behave as a battery; that is, to absorb and release energy in a controllable fashion up to fixed and predetermined limits on volume, charge rate and discharge rate. We derive explicit bounds on the battery capacity that can be offered, and show that there is a fundamental trade-off between the abilities of collective load to absorb and release energy at high aggregate rates. Finally, we introduce a new class of dynamic priority-driven feedback policies that balance these abilities, and characterize the batteries that they can emulate.

## I. INTRODUCTION

The power systems is unique in that generation must meet demand at all times and in the face of uncertainty. To meet this requirement, the system operator relies on balancing reserves, which are often provided by gas power plants or fossil fuel based spinning generators. However, following the recent and rapid developments in communication and metering technology, flexible electricity loads have become viable candidates for providing low-cost, clean and fast-acting balancing services to the power system [1], [2].

Many methods have been suggested for providing demand-based balancing support. An early example can be found in [3], where dispersed loads are equipped with governor-like controllers that respond to changes in system frequency. More recent approaches focus on controlling the aggregate consumption of a collection of loads. This includes work on thermostatically controlled loads [4]–[6], HVAC systems [7], electrical vehicle charging [8], [9], and residential pool pumps [10]. For a comprehensive survey of demand-side balancing support we refer to [1].

Despite the large interest in demand-side balancing services, few attempts have been made to quantify their *capacity*. By capacity we mean the set of aggregate power adjustments that can be tolerated by a collection of loads without causing them unacceptable disruption. The focus of most methods in the literature is not on designing power scheduling policies that are robust to a specific set of power adjustments, but rather, on tracking an arbitrary trajectory. A drawback is that additional effort is required to determine the resulting capacity; for instance, by using the method in [11]. A second and more severe limitation is that the associated capacity under such tracking policies is fixed. As we show in Section IV, the capacity of flexible demand may be subject to substantial trade-offs, and can be adapted to changes in power system operating conditions.

In this paper, we develop scheduling policies that use the flexibility of a collection of deferrable energy loads (i.e., energy demand that can be postponed up to a deadline) to directly *emulate* a battery, which is a useful and well-understood balancing resource. We define a battery in terms of three parameters: the volume of energy that it can store, its maximum charge rate, and its maximum discharge rate. To emulate a battery, the scheduling policy must be able to satisfy the energy requirements of all loads in the collection under any *a priori unknown* supply side fluctuations that are compatible with the charge/discharge behavior of a battery. Our goal is to characterize the batteries that can be emulated and to identify the associated power scheduling policies. To provide insightful answers, we study this question for a collection of homogeneous loads with periodic arrivals.

Our contributions are as follows. We provide upper bounds on the capacity of the batteries that can be emulated in terms of the load data. The bounds show that there are substantial trade-offs between battery parameters. We then show that these trade-offs occur because the initial state (the energy levels of active loads) required to track large positive deviations in aggregate power supply is different than the state required to follow large negative deviations. Since the dynamics of the loads prevent instantaneous transitions between different states, and since the aggregate power adjustment is not known beforehand, there is a fundamental trade-off between the abilities of the collective load to absorb and release energy at high aggregate rates. This trade-off is not an artifact of our homogeneous load model, and is expected to occur in far more general settings. Finally, based on insights from the trade-off analysis we develop a novel class of priority-driven dynamic feedback policies, termed *mixed-slack policies*, and quantify their performance in terms of the battery capacity that they can provide.

Battery models have previously been used in the literature to quantify the aggregate capacity of a col-lection of flexible loads. Here, we let $\mathbb{W}$ denote the aggregate capacity (set of feasible power adjustments) of a load collection and use $\mathbb{B}$ for the capacity of batteries (set of feasible charging trajectories). For instance, in [12], data was used to identify the parameters of a battery such that $\mathbb{B}$ contains $\mathbb{W}$ for a collection of thermostatically controlled loads (TCLs) (i.e., loads whose temperature may fluctuate within a deadband). In [6], the authors give analytical expressions for the battery parameters that satisfy $\mathbb{W} \subset \mathbb{B}$ for a collection of TCLs. An analogous result was presented for deferrable loads in [13], where it is also shown that if all loads are able to consume power at an infinite rate, then the *earliest-deadline-first* policy [14] attains $\mathbb{W} = \mathbb{B}$. Here, we are interested in the batteries that can be emulated by a load collection, i.e, batteries for which $\mathbb{B} \subset \mathbb{W}$. In Section IV, we derive necessary conditions on the parameters of such batteries, which imply that, in general, emulable batteries must be substantially smaller than batteries for which $\mathbb{W} \subset \mathbb{B}$.

In [6], the authors also derive sufficient conditions for $\mathbb{B} \subset \mathbb{W}$ and present a class of linear feedforward policies that emulate the corresponding batteries. This is closely related to our results Section V, where we develop a class of priority-driven feedback policies that can emulate all batteries that satisfy the conditions in Theorem 2. A key difference is that, while a deferrable load is only able to provide temporary energy storage, each thermostatically controlled load can act as a battery on its own. Hence, the problem in [6] is to aggregate a collection of heterogeneous batteries into a larger battery, and in our work it is to aggregate temporary and overlapping storage capacities into a permanent battery. Another difference is that the bounds in [6] are derived under a fixed initial state. Our results show that, at least for deferrable loads, this is restrictive, because different states are required to absorb and release energy at high aggregate rates. A linear programming approach for extracting an emulable battery from a finite collection of heterogeneous deferrable loads was recently proposed in [15]. The method speeds up the computation by using a lift-and-project approach that reduces the number of inequality constraints in the optimization problem. Batteries that satisfy $\mathbb{B} \subset \mathbb{W}$ were also considered in [11], where an empirical method was developed to identify the battery parameters from data. The method is not restricted to a specific policy nor a particular load model.

Several authors consider scheduling of deferrable loads with hard aggregate consumption constraints. For instance, in [16], the authors consider a finite horizon problem where a collection of deferrable loads are subject to a constant aggregate consumption limit, and derive a characterization of the set of all feasible policies. A similar formulation is considered in [17], except that the aggregate power consumption limit varies and is unknown *a priori*. An important implication of this uncertainty is that causal policies fail to take full advantage of the available flexibility. The authors then run simulations to compare the performance of some well-known policies from the literature on processor time allocation, including *least-laxity-first* [18]. In our scheduling problem, the aggregate consumption bound is replaced by a requirement to perfectly track all supply-side fluctuations that are compatible with the charge and discharge pattern of a battery. Interestingly, the least-laxity-first policy emerges as the policy that is most robust to supply shortages, but at the expense of its ability to absorb surpluses.

An initial version of this paper appeared in [19].

*Notation*

For two vectors $x, y \in \mathbb{R}^3$, we say that $x \leq y$ if $x_i \leq y_i$ for all $i = 1, 2, 3$, and that $x < y$ if, in addition, $x_i < y_i$ for some $i$. We define the saturation operator as $[x]_a^b = \max(a, \min(x, b))$.

## II. DEFERRABLE ENERGY CONSUMPTION

A single deferrable energy load is characterized by an arrival time $\tau \in \mathbb{R}$, an energy demand, $E$, a time period, $T$, in which the demand must be filled, and a limit, $\overline{P}$ on its maximum power consumption. The energy consumed at time $t$, by a load with arrival time $\tau$ is denoted

$$e_\tau(t) = \int_{-\infty}^t p_\tau(\theta) d\theta,$$

where $p_\tau$ is the corresponding power consumption. The load admits any power consumption trajectory that satisfies its requirements, that is:

$$e_\tau(t) = 0, \text{ if } t \in (-\infty, \tau] \quad (1a)$$
$$e_\tau(t) = E \text{ if } t \in [\tau + T, \infty) \quad (1b)$$
$$0 \leq p(t) \leq \overline{P} \text{ for all } t \in \mathbb{R} \quad (1c)$$

We define the *nominal power consumption* of the load as

$$p_\tau^{\text{nom}}(t) = \begin{cases} P_0 & t \in [\tau, \tau + T] \\ 0 & \text{otherwise} \end{cases},$$

where $P_0 = E/T$. The flexibility of the load is due to its ability to deviate from nominal consumption. Figure 1 illustrates four possible energy trajectories for a flexible load.

Now consider a collection of deferrable loads that are indexed by their arrival times. We adopt the following assumptions

- Arrival times are deterministic and uniform with rate $\lambda > 0$, i.e., $\tau \in \mathbb{A}_\lambda = \{\ldots, -1/\lambda, 0, 1/\lambda, \ldots,\}$.
- All loads are identical, meaning that they have the same $E$, $T$, and $\overline{P}$.

We wish to use the aggregate flexibility of the loads to comply with an *a priori* unknown and exogenous adjustment, $w$, in the aggregate power consumption.

That is, we wish to select power consumptions, $p_\tau$, that satisfy (1) and

$$\sum_{\tau \in \mathbb{A}_\lambda} p_\tau(t) = \sum_{\tau \in \mathbb{A}_\lambda} p_\tau^{\text{nom}}(t) + w(t). \quad (2)$$

A scheduling policy, $\mu : w \to p$, decides how the aggregate adjustment $w$ should be allocated over the loads. We let $p_\tau = \mu_\tau(w)$ denote the power allocated to load $\tau$. Since $w$ is uncertain, we restrict our attention to *causal* policies, that is, $\mu$ has no *a priori* knowledge of $w$. The *aggregate capacity* or *flexibility* of the loads under a given $\mu$ is defined as

$$\mathbb{W}_\lambda(\mu) = \{w : p_\tau = \mu_\tau(w) \text{ saisfies (1) and (2)}\}$$

To simplify the analysis, we will assume that the arrival rate $\lambda = \infty$. While a generalization of the results in this section to finite arrival rates can be obtained by using essentially the same proof techniques, it makes the analysis more cumbersome and does not add any significant understanding.

To account for infinite arrival rates, we introduce a modified notion of aggregate capacity as follows. By $w \in \alpha \mathbb{W}_\lambda$, where $\alpha$ is a scalar, we mean that $w = \alpha w'$ for some $w' \in \mathbb{W}_\lambda$. We define

$$\mathbb{W}(\mu) = \lim_{\lambda \to \infty} \frac{1}{T\lambda} \mathbb{W}_\lambda(\mu)$$
$$= \left\{ w : \frac{1}{T} \int_\mathbb{R} \mu_\tau(w) d\tau = \frac{1}{T} \int_\mathbb{R} p_\tau^{\text{nom}} d\tau + w, \right.$$
$$\left. \mu_\tau(w) \text{ satisfies (1)} \right\} \quad (3)$$

For large $\lambda$, $T\lambda$ is an accurate estimate of the number of loads that have entered their active consumption phase. In this case,

$$\mathbb{W}_\lambda(\mu) \approx T\lambda \mathbb{W}(\mu) \quad (4)$$

and the set $\mathbb{W}(\mu)$ can be interpreted as the capacity of the average active load.

Let $x_\sigma(t) = e_{t-\sigma}(t)$ denote the energy of the load that arrived $\sigma$ seconds ago. Since it is only necessary to keep track of the energy accumulated by active loads, the *energy allocation*, $x_\sigma, \sigma \in [0, T]$, is a minimal state representation.

*Lemma 1:* The state evolves according to

$$x_{\sigma+h}(t+h) = x_\sigma(t) + \int_t^{t+h} p_{t-\sigma}(\theta) d\theta, \quad (5)$$

for $h \geq 0$. Moreover, if $\partial x_\sigma(t)/\partial \sigma$ exists, then

$$\frac{\partial x_\sigma(t)}{\partial t} + \frac{\partial x_\sigma(t)}{\partial \sigma} = p_{t-\sigma}(t). \quad (6)$$

$\triangledown$

From (1) it follows that

$$\underline{x}_\sigma \leq x_\sigma(t) \leq \overline{x}_\sigma, \quad (7)$$

where

$$\overline{x}_\sigma = \left[\overline{P}\sigma\right]_0^E \quad \text{and} \quad \underline{x}_\sigma = \left[\overline{P}(\sigma - (1 - \frac{P_0}{\overline{P}})T)\right]_0^E.$$

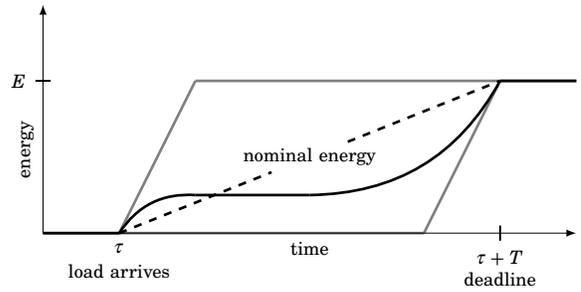

Fig. 1: Feasible energy trajectories for a deferrable load. The upper and lower gray curves correspond to the trajectories that fill the energy demand in the shortest time possible and postpone the consumption for as long as possible, respectively. The dashed line shows the nominal energy trajectory.

These bounds are illustrated by the gray lines in Figure 3 and are attained when all loads fill their demands as fast as possible, or defer their consumption for as long as possible.

The (average) aggregate energy that is *stored* by the loads (in addition to their nominal energy) is denoted by

$$x_{\text{avg}}(t) = \frac{1}{T} \left( \int_0^T x_\sigma(t) d\sigma - \int_0^T x_\sigma^{\text{nom}} d\sigma \right), \quad (8)$$

where $x_\sigma^{\text{nom}} = [P_0 \sigma]_0^E$.

## III. BATTERY EMULATION

Our objective is to develop scheduling policies that allow a collection of deferrable loads to act as battery. An ideal battery is characterized by a volume, $C$, charge rate $\overline{W}$, and discharge rate $\underline{W}$. The exogenous power trajectories that can be absorbed by a battery with parameters $\phi = (C \ \overline{W} \ \underline{W})$ are given by

$$\mathbb{B}(\phi) = \left\{ w : -\underline{W} \leq w(t) \leq \overline{W}, \ -\frac{C}{2} \leq \chi(t) \leq \frac{C}{2}, \right.$$
$$\left. \chi(t) = \int_{-\infty}^t w(\theta) d\theta, \ w(t) = 0 \text{ for } t < 0 \right\} \quad (9)$$

*Definition 1 (Realizable battery):* We say that a policy $\mu$ *realizes* the battery $\mathbb{B}(\phi)$ if $\mathbb{B}(\phi) \subset \mathbb{W}(\mu)$. The battery $\mathbb{B}(\phi)$ is called *realizable*, written $\mathbb{B}(\phi) \subset \mathbb{W}$, if there is *some* causal policy that realizes it.

*Definition 2 (Maximal battery):* We say that a battery $\mathbb{B}(\phi) \subset \mathbb{W}$ is *maximal* if there is no strictly larger realizable battery, that is, there is no $\phi'$ such that $\mathbb{B}(\phi') \subset \mathbb{W}$ and $\mathbb{B}(\phi') \supsetneq \mathbb{B}(\phi)$.

Given load parameters $E$, $T$, $\overline{P}$, our goal is to
- characterize the set of maximal batteries
- develop scheduling policies that realize them

*Remark 1 (Initialization phase):* As we show in Section IV, different initial energy allocations may be required for emulating different batteries. This means that, before starting the battery service, an initialization phase is required during which $\mu$ attains the required $x(0)$. For notational convenience, the initialization period is set to $(-\infty, 0)$. It is worth mentioning that the policies we develop in Section V correctly initialize the loads in finite time. ▽

## IV. BATTERY PARAMETER TRADE-OFFS

In this section, we show that there are fundamental trade-offs in the battery parameters that can be attained, and explain why this is the case.

*Proposition 1:* Let $\phi_{\max} = \begin{bmatrix} C_{\max} & \overline{W}_{\max} & \underline{W}_{\max} \end{bmatrix}$, where

$$C_{\max} = E(1 - \tfrac{P_0}{\overline{P}}) \qquad \overline{W}_{\max} = \overline{P} - P_0 \qquad \underline{W}_{\max} = P_0. \tag{10}$$

Then $\mathbb{B}(\phi_{\max})$ is the smallest set that contains all $\mathbb{B}(\phi) \subset \mathbb{W}$. ▽

A consequence of Proposition 1 is that if there is no trade-off between battery parameters, that is, if there is a single largest realizable battery, it must be $\mathbb{B}(\phi_{\max})$. The next result shows that, in practice, this is not the case.

*Proposition 2:* $\mathbb{B}(\phi_{\max}) \subset \mathbb{W}$ if and only if $\overline{P} = \infty$. ▽

*Remark 2 (Connection to previous work):* When all loads are identical, the limits presented in [13, Section IV-B] imply that $\mathbb{B}(\phi_{\max})$ does contains all realizable batteries, but do not establish tightness of this bound. Also, it follows from the results in [13, Section III] that if the loads are able to consume power at an infinite rate, then it is possible to emulate $\mathbb{B}(\phi_{\max})$. This covers sufficiency in Proposition 2 but not necessity. ▽

If we disregard the trivial case $\mathbb{B}(\phi) = \{0\}$ (for instance this includes $\phi = (\infty\ 0\ 0)$) then $\phi \leq \phi_{\max}$ is a necessary condition for $\mathbb{B}(\phi) \subset \mathbb{W}$. Henceforth, we will assume:

$\mathcal{A}_1$: $\overline{P} < \infty$.

$\mathcal{A}_2$: $\phi \leq \phi_{\max}$

*Theorem 1:* Suppose $\mathcal{A}_{1\text{-}2}$ hold and set

$$c = C/C_{\max}, \quad \overline{w} = \overline{W}/\overline{W}_{\max}, \quad \underline{w} = \underline{W}/\underline{W}_{\max}. \tag{11}$$

Then $\mathbb{B}(\phi) \subset \mathbb{W}$ only if

$$(\overline{w} + \underline{w} - c)^2 \leq 4\overline{w}\underline{w}(1-c), \tag{12}$$

whenever $\overline{w}, \underline{w} \geq 1 - c$ and $\overline{w} + \underline{w} \geq c$. ▽

Theorem 1 provides upper bounds on realizable batteries. These bounds are depicted in Figure 2 for different energy storage capacities and show that any realizable $\mathbb{B}(\phi)$ must be substantially smaller that $\mathbb{B}(\phi_{\max})$. In particular, if we try to realize a battery where any two of the battery parameters are set at their individual bounds, the third must be zero. For instance, if we insist on a battery that is able to both charge at rate $\overline{W}_{\max}$ and discharge at $\underline{W}_{\max}$, then the resulting energy storage capacity is zero. It will be shown in Theorem 2 that the bounds in Theorem 1 are optimistic.

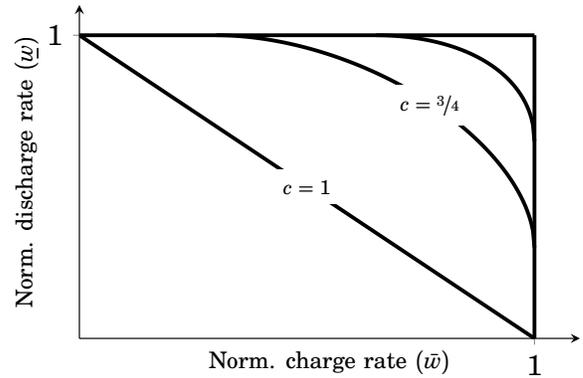

Fig. 2: Limits on normalized battery charge and and discharge rates that can be attained. The different curves correspond to normalized storage capacity $c = \{0, 0.33, 0.75, 1\}$.

*Remark 3 (Non-uniform rates):* A reason behind the substantial trade-offs in Theorem 1 is the requirement to guarantee *uniform* charge/discharge rates over all energy levels in $[-C/2, C/2]$. To better utilize the flexibility of the loads, one could replace $\mathbb{B}(\phi)$ with a different type of storage device where $\overline{W}$ and $\underline{W}$ depend on $\chi$. How to choose this dependence in order to best utilize the available flexibility is an interesting research direction.

Next, we introduce two notions that will be central in explaining the trade-offs between battery parameters, and in designing policies that emulate batteries in the next section. We define the *charge slack* and *discharge slack* of a load that arrived $\sigma$ seconds ago as

$$\overline{s}_\sigma(t) = (E - x_\sigma(t))/\overline{P} \tag{13}$$
$$\underline{s}_\sigma(t) = T - \sigma - \overline{s}_\sigma(t) \tag{14}$$

The charge slack is the longest period that the load can maintain maximum consumption rate, $\overline{P}$. It constitutes a measure of the loads ability to *absorb* energy. Similarly, the discharge slack is the longest duration the load can maintain minimum, i.e., zero, consumption rate, and quantifies its ability to *release* energy. To emulate a battery which is able to both absorb and release large volumes of energy at high rates, it is necessary to maintain high charge and discharge slacks across all loads. However, this is not possible for two reasons. First, $\overline{s}_\sigma + \underline{s}_\sigma = T - \sigma$, and cannot be modified via the power consumption. This means that an increase in $\overline{s}_\sigma$ implies a decrease in $\underline{s}_\sigma$ and vice versa. Second, the requirement that $x_{\text{avg}}(t) = \chi(t)$ implies that $\int \overline{s}_\sigma$ and $\int \underline{s}_\sigma$ are completely determined by $\chi$. Hence, an increase in, say, $\overline{s}_\sigma$ must come at the expense of decreasing some other $\overline{s}_{\sigma'}$.

We now investigate the slack requirements of two illustrative batteries that satisfy (12).

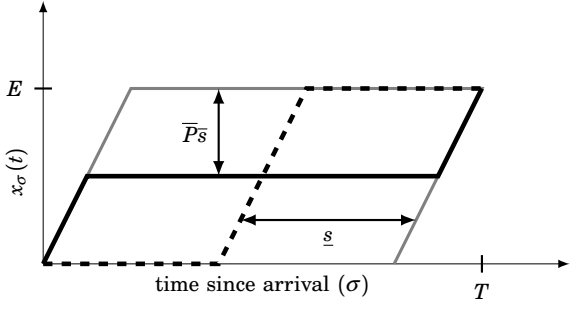

Fig. 3: Depiction of energy allocations, $x_\sigma(t)$, that satisfy (15) (solid) and (16) (dashed) for a fixed $t$. In both cases $x_{\text{avg}}(t) = 0$. The gray lines show the maximum and minimum attainable energy levels $\overline{x}$ and $\underline{x}$.

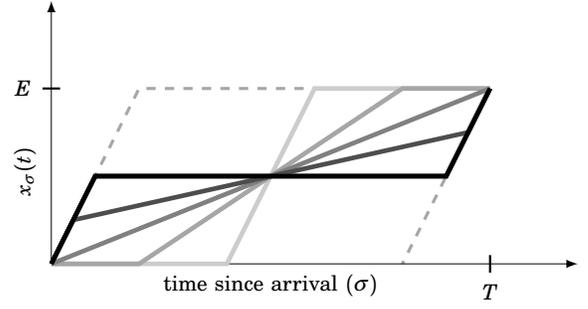

Fig. 4: The solid lines illustrate the equilibrium state, $x(t)$, under $\mu^\eta$ for $\eta = \{0, \frac{P_0}{2}, \frac{P_0}{\overline{P}}, \frac{2P_0}{\overline{P}+P_0}, 1\}$. The stored energy $x_{\text{avg}}(t) = 0$ in all cases. Darker lines correspond to smaller $\eta$. Equilibrium states corresponding to higher values $\eta$ enable the loads to release energy at higher aggregate rates, while smaller $\eta$ allow for faster energy absorption. The gray dashed lines depict $\overline{x}$ and $\underline{x}$.

*Proposition 3:* Let $\overline{\phi} = (C_{\max}, \overline{W}_{\max}, 0)$ and suppose that $\mathbb{B}(\overline{\phi}) \subset \mathbb{W}(\mu)$. Let $w \in \mathbb{B}(\overline{\phi})$ and set $p = \mu(w)$. Then the following implication holds

$$\overline{s}_\sigma(t) > \overline{s}_{\sigma'}(t) \implies x_\sigma(t) = \overline{x}_\sigma \text{ or } x_{\sigma'}(t) = \underline{x}_{\sigma'} \quad (15)$$

for any $\sigma, \sigma' \in [0, T]$ and $t \in \mathbb{R}$. ▽

By Theorem 1, the battery $\mathbb{B}(\overline{\phi})$ is able to absorb large volumes of energy faster than any other realizable battery. Proposition 3 states that the state trajectory under the policy that realizes $\mathbb{B}(\overline{\phi})$ must be as even as possible in terms of its charge-slack profile. Similarly, the next result shows that the policy that realizes the battery that is best suited to release large volumes of energy strives to equalize the discharge slacks.

*Proposition 4:* Let $\underline{\phi} = (C_{\max}, 0, \underline{W}_{\max})$ and suppose that $\mathbb{B}(\underline{\phi}) \subset \mathbb{W}(\mu)$. Let $w \in \mathbb{B}(\underline{\phi})$ and set $p = \mu(w)$. Then the following implication holds

$$\underline{s}_\sigma(t) > \underline{s}_{\sigma'}(t) \implies x_\sigma(t) = \underline{x}_\sigma \text{ or } x_{\sigma'}(t) = \overline{x}_{\sigma'} \quad (16)$$

for any $\sigma, \sigma' \in [0, T]$ and $t \in \mathbb{R}$. ▽

The energy allocations that satisfy (15) and (16) are shown in Figure 3. To understand why there is a trade-off between the batteries that can be emulated, suppose that $\mathbb{B}(\phi_{\max}) \subset \mathbb{W}(\mu)$ for some causal policy $\mu$, which, by Proposition 1, would be the case if there was no trade-off. Since $w$ is not known in advance, at time $t$, $\mu$ must be able to manage both of the following two outcomes: absorb energy at an aggregate rate $\overline{W}_{\max}$ until $x_{\text{avg}}(t') = C_{\max}/2$, and release energy at an aggregate rate $\underline{W}_{\max}$ until $x_{\text{avg}}(t'') = -C_{\max}/2$, for some $t', t'' \geq t$. The first scenario requires *all* loads in their active consumption phase to consume at the maximum rate, $\overline{P}$, until *all* loads have reached their highest attainable energy level. This is only possible if $x(t)$ satisfies (15). The second scenario requires *all* loads in their active consumption phase to stop consumption until *all* loads have reached their lowest attainable energy level. This is only possible if $x(t)$ satisfies (16). Since it is not possible to instantaneously transition between energy allocations that satisfy (15) and (16), some battery capacity must be sacrificed. If we reduce $\overline{W}$ ($\underline{W}$) we relax the requirement that all of the active loads must be able to maintain their maximum (minimum) consumption rate for the duration it takes the battery to fully charge (discharge). If we reduce $C$, we relax the requirement that all loads must have reached their highest (lowest) attainable energy levels for the battery to be considered fully charged (discharged).

The discussion above shows that there is a fundamental trade-off between the abilities of a collection of deferrable loads to absorb and release energy at high rates. This trade-off is not an artifact of our homogeneous and deterministic load model. It is a consequence of 1) the energy allocations required for absorbing and releasing energy at high rates have different slack profiles, 2) the dynamics of the loads prevent instantaneous transitions between these allocations, and 3) due to the uncertainty in the reference trajectory, $w$, it is not possible to infer the required energy allocation ahead of time. Since the definitions of charge and discharge slack naturally extend to non-identical loads, the identified trade-off occurs for a wide range of collections of heterogeneous and uncertain flexible loads as well.

## V. SCHEDULING AND REALIZABLE BATTERIES

In this section we develop a family of heuristic scheduling policies that can realize a wide range of batteries. The basis for our heuristic is the observation that the policies in Proposition 3 and Proposition 4 strive to *equalize* the charge and discharge slack, respectively. This allows us to express these policies as priority-driven feedback policies. At each time $t$, the policy in Proposition 4 allocates as much power as possible, i.e., without violating (1), to the loads with the smallest $\underline{s}_\sigma(t)$, while the policy in Proposition 3 allocates as much power as possible to the loads with the largest $\bar{s}_\sigma(t)$, or equivalently, the smallest $\overline{E}/\overline{P} - \bar{s}_\sigma(t) = x_\sigma(t)/\overline{P}$. To obtain a policy which is more balanced in terms of charge/discharge rates we pick $\eta \in [0, 1]$ and introduce the *mixed slack*

$$s_\sigma^\eta = \eta \underline{s}_\sigma + (1-\eta)(E/\overline{P} - \bar{s}_\sigma). \tag{17}$$

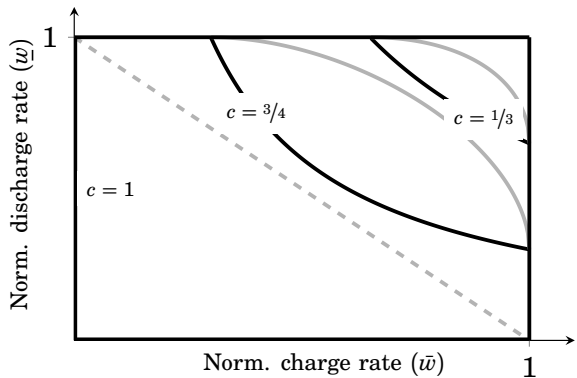

Fig. 5: Attainable trade-offs in battery capacity with mixed-slack policies. The different curves correspond to $c = \{0, 0.33, 0.75, 1\}$. The gray curves show the upper bounds from Figure 2. For $c = 1$ condition (18) is an exact characterization, which means that the bound indicated by the dashed gray curve is optimistic.

*Definition 3 (Mixed-slack policy):* The policy that allocates as much power as possible to the loads with the smallest $s^\eta$ is denoted $\mu^\eta$. We refer to these type of policies as *mixed-slack policies* (MSPs).

Some equilibrium states under $\mu^\eta$ for different $\eta$ are illustrated in Figure 4. The batteries that can be realized by MSPs are given by the next theorem.

*Theorem 2:* Suppose $\mathcal{A}_{\text{1-2}}$ hold and that

$$\overline{w}\underline{w} + c \leq 1, \tag{18}$$

where $c$, $\overline{w}$, and $\underline{w}$ are defined in (11). Then $\mathbb{B}(\phi) \subset \mathbb{W}(\mu^\eta)$ where $\eta = 1/(1 + \overline{W}/\underline{W})$. Moreover, if either $c = 1$, $\overline{w} = 1$, or $\underline{w} = 1$, then (18) is also necessary for $\mathbb{B}(\phi) \subset \mathbb{W}$. ▽

Figure 5 illustrates the battery parameter trade-offs under MSPs. In general, there is a gap between the upper bounds from Theorem 1 and the lower bounds in Theorem 2. This means that all maximal batteries might not be realized by MSPs. However, if we insist on setting any of the battery parameters at its upper bound, then Theorem 2 provides an exact characterization of the attainable trade-off between the remaining two parameters,. A consequence of this is that the upper bound for $c = 1$ is optimistic. In fact, when emulating a battery with the maximum attainable volume, either the charge or discharge rate must be zero.

*Remark 4 (MSPs in more general settings):* MSPs are feedback policies and the definitions of charge and discharge slack is valid for non-identical loads as well. This implies that MSPs may also be useful for scheduling a wider range heterogeneous load collections.

*Remark 5 (Relation to scheduling literature):* In the scheduling literature, the MSP $\mu^1$ is known as the *least-laxity first* (LLF) policy [18]. Recently, the LLF policy was used in [17] to schedule a collection of deferrable energy consumers subject to uncertain supply side fluctuations. Given that we now know that LLF is ideal for withstanding negative fluctuations, but less robust to positive fluctuations, it would be interesting to revisit this study and include more balanced MSPs.

## VI. NUMERICAL EXAMPLE

Consider an example, where a *load aggregator* manages the energy consumption of a pool of electrical vehicles. We assume that each vehicle requires $E = 60$ kWh of energy within a period of $T = 10$ h. The nominal charge rate for each vehicle is $P_0 = 6$ kW, and the maximum charge is $\overline{P} = 18$ kW. Further, we assume that vehicles plug in periodically at a rate of $\lambda = 10$ vehicles per hour. This implies that at any point in time, a number of $T\lambda = 100$ vehicles are in their active consumption phase and that the aggregate nominal consumption is 600 kW.

While our example is rather stylized, the parameters above are indicative of real electrical vehicle batteries. Therefore, the batteries that are presented below serve as an estimate of the battery capacity that can be realized by managing a substantial number of electrical vehicles.

Plugging the load parameters into the expressions in Proposition 1, gives $C_{\max} = 40$ kWh, $\overline{W}_{\max} = 12$ kW, and $\underline{W}_{\max} = 6$ kW. We give three examples of batteries that can be realized by the load aggregator.

Let $\phi_A = \left(\frac{3C_{\max}}{4} \quad \frac{\overline{W}_{\max}}{4} \quad \underline{W}_{\max}\right)$. By Theorem 2, $\mathbb{B}(\phi_A) \subset \mathbb{W}(\mu_A)$, where $\mu_A = \mu^{2/3}$. Setting $\Phi_A = T\lambda \phi_A$ and using (4) we have

$$\mathbb{B}(\Phi_A) = T\lambda \mathbb{B}(\phi_A) \subset T\lambda \mathbb{W}(\mu_A) \approx \mathbb{W}_\lambda(\mu_A).$$

We conclude that $\mu_A$ can (approximately) emulate the battery $\mathbb{B}(\Phi_A)$ with volume, charge rate, and discharge rate

$$C_A = 3 \text{ MWh} \quad \overline{W}_A = 300 \text{ kW} \quad \underline{W}_A = 600 \text{ kW}$$

Let $\Phi_B = T\lambda(\frac{C_{\max}}{4} \quad \frac{3\overline{W}_{\max}}{4} \quad \underline{W}_{\max})$ and $\Phi'_B = T\lambda(\frac{11}{12}C_{\max} \quad \frac{1}{4}\overline{W}_{max} \quad \frac{1}{3}\underline{W}_{\max})$. Using similar steps as above we conclude that $\mu_B$ can (approximately) emulate the battery $\mathbb{B}(\Phi_B)$ with

$$C_B = 1 \text{ MWh} \quad \overline{W}_B = 900 \text{ kW} \quad \underline{W}_B = 600 \text{ kW}$$

as well as the battery $\mathbb{B}(\Phi'_B)$ with

$$C'_B = 3.67 \text{ MWh} \quad \overline{W}'_B = 300 \text{ kW} \quad \underline{W}'_B = 200 \text{ kW}$$

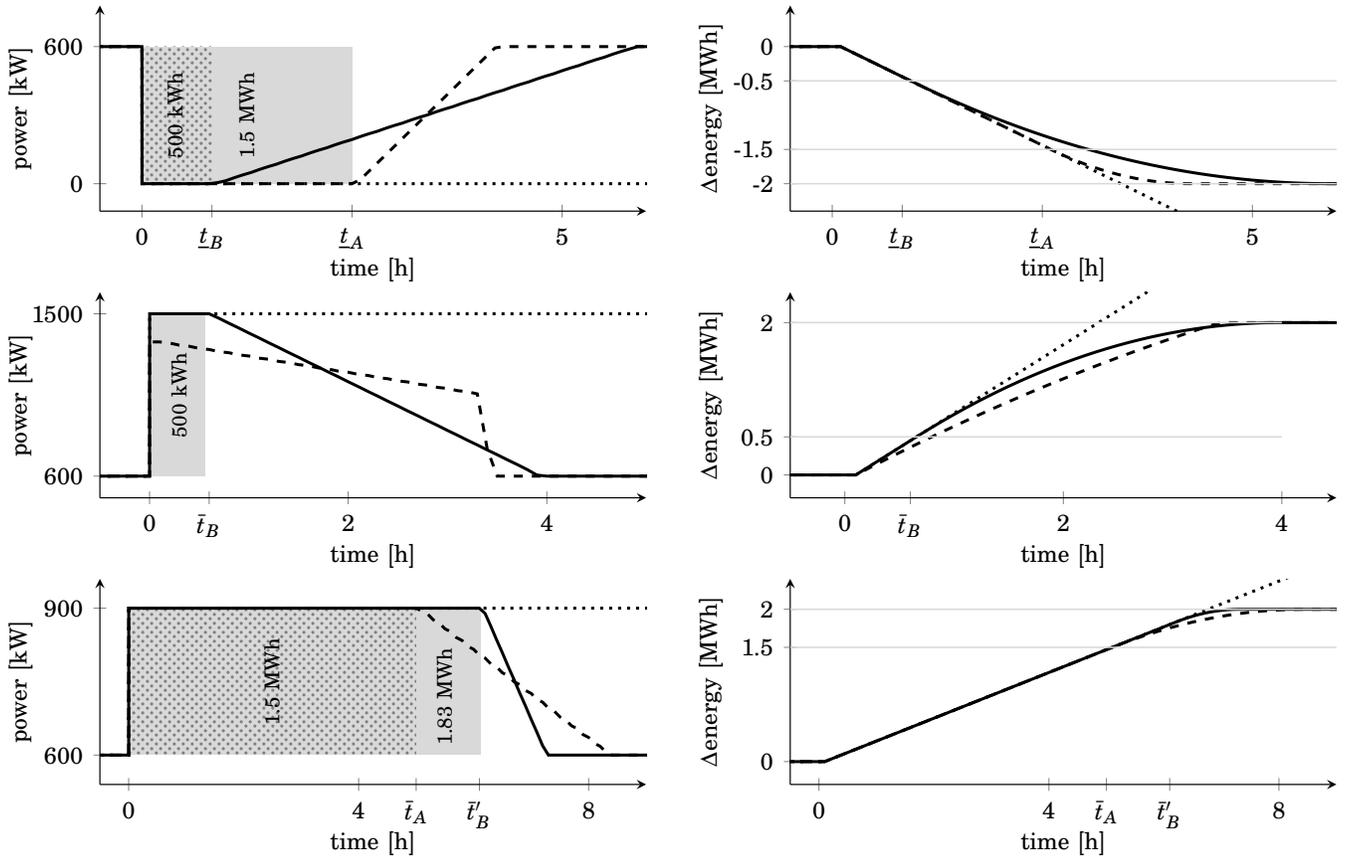

Fig. 6: Absolute aggregate power consumption and aggregate stored energy under $\mu_A$ (dashed) and $\mu_B$ (solid) in response to a set-point change (dotted). The upper plots shows the response to a request of releasing energy at a rate of 600 kW. Policy $\mu_B$ fails to maintain the discharge rate after releasing $C_B/2 = 500$ kWh, while $\mu_A$ manages to maintain the desired rate until $C_A/2 = 1.5$ MWh. The middle plots show the response to a power set-point of $\overline{W}_B = 900$ kW. Policy $\mu_A$ cannot provide the desired charge rate, while $\mu_B$ manages to maintain it until having absorbed $C_B/2 = 500$ kWh of energy. The lower plots show the response to a power set-point of $\overline{W}_A = 300$ kW. The policies $\mu_A$ and $\mu_B$ track the set-point until they have absorbed $C_A/2 = 1.5$ MWh and $C'_B/2$ MWh, respectively.

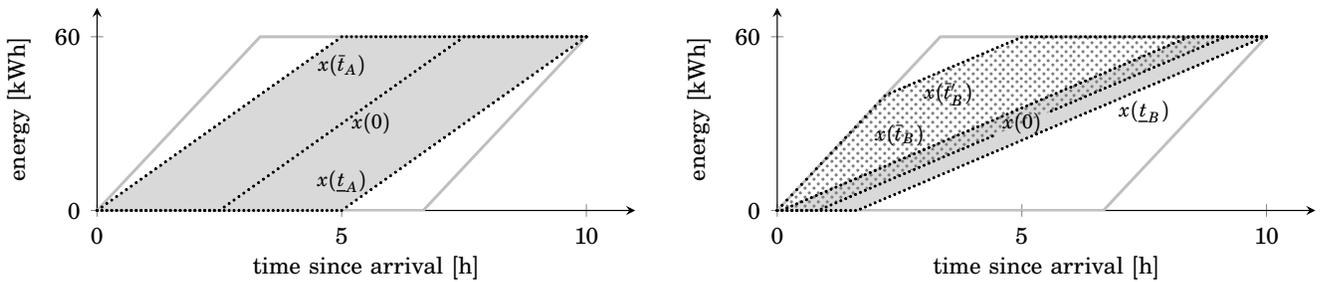

Fig. 7: Energy allocation, $x_\sigma$, at times when $\mu_A$ (left) and $\mu_B$ (right) fail to track the different power set-points. Each mark represents the energy consumed by an active load and the gray lines show the energy bounds $\overline{x}$ and $\underline{x}$. The gray area in the left and right plots equals $C_A$ and $C_B$, respectively, and the dotted area equals $C'_B/2$. The policy $\mu_A$ fails to track $\underline{W}_A$ at time $\underline{t}_A$, which is when the first active load has exhausted its discharge slack and can no longer defer its consumption. The policy $\mu_B$ fails to track $W_A$ at time $\underline{t}_B$ for the same reason. Similarly, $\mu_A$ and $\mu_B$ fail to track positive set-points when sufficiently many loads have exhausted their charge slacks.

Note that $\mathbb{B}(\Phi_A)$ and $\mathbb{B}(\Phi_B)$ are (approximately) maximal batteries and that $\mathbb{B}(\Phi_B)$ and $\mathbb{B}(\Phi'_B)$ are realized by the same policy.

Next, we investigate the ability of $\mu_A$ and $\mu_B$ to absorb and release energy by simulating the response in aggregate power consumption to different set-point changes. See Figure 6. The initial energy allocation as well as the energy allocations at the time the policies fail to track the set-point are illustrated in Figure 7. In each simulation, the initial state, $x(0)$, is at equilibrium, and the initial stored aggregate energy is zero (which means that $\mathbb{B}(\Phi_A)$, $\mathbb{B}(\Phi_B)$ and $\mathbb{B}(\Phi'_B)$ come half full). We use a time step of $1/\lambda$ h for updating the energy levels of the loads.

In the simulations, there are small discrepancies between the times that $\mu_A$ and $\mu_B$ are expected to fail in tracking the set-point according to $\mathbb{B}(\Phi_A)$, $\mathbb{B}(\Phi_B)$ and $\mathbb{B}(\Phi'_B)$ and the times that they actually do fail. These discrepancies are due to the finite arrival rate. We ran several simulations with different parameters and the the discrepancies in time of failure were always less than $1/\lambda$.

## VII. SUMMARY

We investigated the possibility of using the flexibility of a collection of deferrable loads to emulate a battery. A battery captures the three most important properties of energy storage: the volume of energy that can be stored, the rate at which it can be absorbed, and the rate at which it can be released.

We derived upper bound on the battery capacity that can be realized, and showed that there is a fundamental trade-off between the abilities of collective load to absorb and release energy at high aggregate rates. By analyzing the scheduling policies that emulate two illustrative batteries, we found that these trade-offs are due to a conflict between maintaining even charge and discharge slacks simultaneously across all active loads. Based on these insights, we developed a novel family of dynamic priority-driven scheduling policies, termed *mixed-slack policies* (MSP). Each MSP assigns priorities based on a unique weighted combination of charge and discharge slacks, and strikes a different trade-off between the ability to absorb and release energy. By characterizing the set of batteries that can be emulated with MSPs, we obtained a lower bound on the capacity of maximal batteries.

Future research includes working towards an exact characterization of the batteries that can be emulated, and emulating other types of devices. Another research direction is to account for stochasticity and heterogeneity in load parameters (arrivals, energy demand, etc.) and characterize the resulting trade-off between *reliability* and capacity.

## APPENDIX

We start by introducing some useful lemmas. The following observation will be used frequently:

*Lemma 2:*
$$\frac{1}{T}\int_0^T (\overline{x}_\sigma - x_\sigma^{\text{nom}})\,d\sigma = \frac{1}{T}\int_0^T (x_\sigma^{\text{nom}} - \underline{x}_\sigma)\,d\sigma = \frac{C_{\max}}{2},$$

where $C_{\max}$ is defined in (10).

*Proof:* This follows immediately from the definitions of $\overline{x}$ and $\underline{x}$ (see Figure 3) and $x_\sigma^{\text{nom}} = P_0\sigma$. ∎

If $w \in \mathbb{B}(\phi)$, the maximum duration the battery can maintain its maximum and minimum rates are

$$t_c(\chi) = \frac{C/2 - \chi}{\overline{W}} \quad \text{and} \quad t_d(\chi) = \frac{C/2 + \chi}{\underline{W}},$$

where the energy level, $\chi$, is defined in (9). The next result gives a necessary condition for $\mathbb{B}(\phi) \subset \mathbb{W}(\mu)$.

*Lemma 3:* Suppose that $\mathbb{B}(\phi) \subset \mathbb{W}(\mu)$ and let $p = \mu(w)$. Then, for each $w \in \mathbb{B}(\phi)$, there are functions $\underline{\delta}_\sigma, \overline{\delta}_\sigma : \mathbb{R} \to [0, \overline{E}]$, such that

$$\int_0^T \underline{\delta}_\sigma(t)d\sigma \leq \frac{T(C_{\max} - C)}{2} \tag{19}$$

$$\int_0^T \overline{\delta}_\sigma(t)d\sigma \leq \frac{T(C_{\max} - C)}{2} \tag{20}$$

$$\overline{z}_\sigma - \overline{\delta}_\sigma \leq x_\sigma \leq \underline{z}_\sigma + \underline{\delta}_\sigma \tag{21}$$

where

$$\overline{z}_\sigma(\chi) = \overline{x}_{\sigma + t_c(\chi)} - \overline{P}t_c(\chi) \quad \text{and} \quad \underline{z}(\chi) = \underline{x}_{\sigma + t_d(\chi)}. \tag{22}$$

*Proof:* Under the assumptions of the lemma, we have $x_{\text{avg}}(t) = \chi(t)$. Since $w$ is not known to $\mu$ a priori, at time $t$, $\mu$ must be able to handle both of the following outcomes

1) $w(\theta) = \overline{W}$ for $\theta \in [t, t + t_c]$
2) $w(\theta) = -\underline{W}$ for $\theta \in [t, t + t_d]$

These outcomes correspond to charging/discharging the battery at maximum rate until it is full/empty. If the outcome is 1), then $x_{\text{avg}}(t + t_c) = \chi(t + t_c) = C/2$. We can express $x(t + t_c) = \overline{x} - \delta$, for some $0 \leq \delta_\sigma \leq \overline{x}_\sigma - \underline{x}_\sigma \leq E$. Using this in (8), yields

$$\frac{C}{2} = \frac{1}{T}\int_0^T (x_\sigma(t + t_c) - x_\sigma^{\text{nom}})\,d\sigma$$

$$= \frac{1}{T}\int_0^T (\overline{x}_\sigma - \delta_\sigma - x_\sigma^{\text{nom}})\,d\sigma = \frac{C_{\max}}{2} - \frac{1}{T}\int_0^T \delta_\sigma d\sigma$$

$$\implies \frac{1}{T}\int_0^T \delta_\sigma d\sigma = \frac{(C_{\max} - C)}{2}$$

where the third equality follows from Lemma 2. From (5) and (1c) we get

$$\overline{x}_{\sigma + t_c} - \delta_{\sigma + t_c} = x_{\sigma + t_c}(t + t_c) \leq x_\sigma(t) + \overline{P}t_c$$

Setting $\overline{\delta}_\sigma(t) = \delta_{\sigma + t_c}$, gives the first inequality in (21). Also, $\overline{\delta}_\sigma \in [0, E]$ and

$$\int_0^T \overline{\delta}_\sigma d\sigma = \int_{t_c}^{T+t_c} \delta_\sigma d\sigma = \int_{t_c}^T \delta_\sigma d\sigma \leq \frac{(C_{\max} - C)T}{2}$$

which establishes (20). The steps to verify (19) and the second inequality in (21) are analogous. ∎

### A. Proof of Lemma 1

We have

$$x_{\sigma+h}(t+h) = e_{t+h-(\sigma+h)}(t+h) = e_{t-\sigma}(t+h)$$

$$= e_{t-\sigma}(t) + \int_t^{t+h} p_{t-\sigma}(\theta)d\theta = x_\sigma(t) + \int_t^{t+h} p_{t-\sigma}(\theta)d\theta$$

To show (6), let $f(t, \tau) = e_\tau(t)$ and $g(t, \sigma) = x_\sigma(t) = f(t, t - \sigma)$. We have $\frac{\partial g}{\partial \sigma} = -\frac{\partial f}{\partial \tau}$. Hence,

$$\frac{\partial g}{\partial t} = \frac{\partial f}{\partial t} + \frac{\partial f}{\partial \tau}\frac{\partial \tau}{\partial t} = \frac{\partial f}{\partial t} + \frac{\partial f}{\partial \tau} = \frac{\partial f}{\partial t} - \frac{\partial g}{\partial \sigma}.$$

### B. Proof of Proposition 1

Let $w \in \mathbb{W}(\mu)$ for some policy $\mu$ and set $p_\tau = \mu_\tau(w)$. Then

$$Tw(t) = \int_\mathbb{R} p_\tau(t) - p_\tau^{\text{nom}}(t)d\tau$$

$$= \int_{t-T}^t p_\tau(t) - P_0 d\tau \leq (\overline{P} - P_0)T$$

and similarly we can show that $w(t) \geq -P_0$. The energy stored by the collection at time $t$ is $\int_0^t w(\theta)d\theta = x_{\text{avg}}(t)$, where $x_{\text{avg}}$ is defined by (8). Using that $x(t) \leq \overline{x}$ we have

$$\int_0^t w(\theta)d\theta \leq \frac{1}{T}\int_0^T (\overline{x}_\sigma - x_\sigma^{\text{nom}})d\sigma = \frac{C_{\max}}{2},$$

where the last equality follows from Lemma 2. Similarly, we can show that $\int_{-\infty}^t w(\theta)d\theta \geq -C_{\max}/2$. Hence, $\mathbb{W}(\mu) \subset \mathbb{B}(\phi_{\max})$ and it follows that $\mathbb{B}(\phi_{\max})$ contains all realizable batteries.

To show that $\mathbb{B}(\phi_{\max})$ is the smallest battery that contains all realizable batteries it suffices to prove the implication

$$\mathbb{B}(\phi) \subset \mathbb{B}(\phi') \text{ for all } \mathbb{B}(\phi) \subset \mathbb{W} \implies \phi' \geq \phi_{\max} \tag{23}$$

By Theorem 2, which is established independently of (23), $\mathbb{B}(\overline{\phi}) \subset \mathbb{W}$ and $\mathbb{B}(\underline{\phi}) \subset \mathbb{W}$, where $\overline{\phi} = \begin{bmatrix} C_{\max} & \overline{W}_{\max} & 0 \end{bmatrix}$ and $\underline{\phi} = \begin{bmatrix} C_{\max} & 0 & \underline{W}_{\max} \end{bmatrix}$. Hence, by assumption, $\mathbb{B}(\overline{\phi}) \subset \mathbb{B}(\phi')$ and $\mathbb{B}(\underline{\phi}) \subset \mathbb{B}(\phi')$. It follows that $\phi' \geq \overline{\phi}$ and $\phi' \geq \underline{\phi}$.

### C. Proof of Proposition 2

Necessity follows from Theorem 1, which is established independently of Proposition 2. For the converse direction, we will show that if $\overline{P} = \infty$, then $\mathbb{B}(\phi_{\max}) \subset \mathbb{W}(\mu)$, where $\mu$ allocates power according to

$$x_\sigma(t) = \begin{cases} \underline{x}_\sigma & \text{if } \sigma \in [0, T - \varsigma(t)] \\ \overline{x}_\sigma & \text{if } \sigma \in (T - \varsigma(t), T] \end{cases} \tag{24}$$

and where $\varsigma(t) = \frac{C_{\max}/2 + \chi(t)}{\overline{W}_{\max}}$. It is straightforward to verify that under (24) we have $x_{\text{avg}}(t) = \chi(t)$ for all $t$, so $p_\tau = \mu_\tau(w)$ satisfies

$$\frac{1}{T}\int_\mathbb{R} \mu_\tau(w)d\tau = \frac{1}{T}\int_\mathbb{R} p_\tau^{\text{nom}}d\tau + w \tag{25}$$

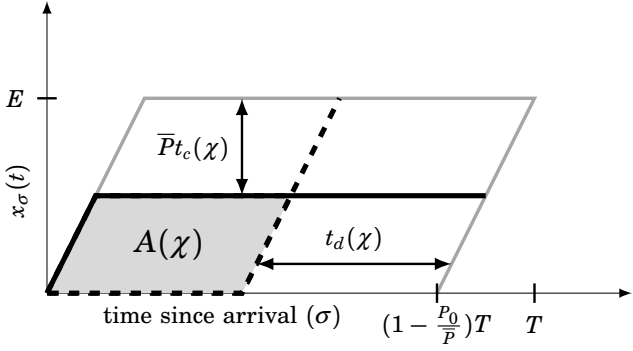

Fig. 8: Illustration of $A(\chi)$ in condition in (27). The solid line shows $\bar{z}_\sigma$ for $\sigma \in [0, T-t_c]$ and the dashed line shows $\underline{z}_\sigma$ for $\sigma \in [0, T-t_d]$.

Also, as long as $\chi(t) \in [-\frac{C_{\max}}{2}, \frac{C_{\max}}{2}]$ we have $\varsigma(t) \in [0,T]$. It follows that $\mu_\tau(w)$ satisfies (1) for every $w \in \mathbb{B}(\phi_{\max})$.

### D. Proof of Theorem 1

By $\mathcal{A}_2$ it is sufficient to show that if $\mathbb{B}(\phi) \subset \mathbb{W}_\infty$ then (12) is satisfied on

$$\mathbb{S} = \{c, \bar{w}, \underline{w} \in [0,1]: \bar{w}, \underline{w} \geq 1-c, \bar{w}+\underline{w} \geq c\}.$$

The only point in $\mathbb{S}$ that corresponds to $\bar{w} = 0$ has $\underline{w} = 1$ and $c = 1$. This point satisfies (12). Similarly, condition (12) is also satisfied when $\underline{w} = 0$. Hence we may restrict our attention to $\bar{w}, \underline{w} > 0$.

It follows from Lemma 3 that for each $\chi \in [-C/2, C/2]$, there are non-negative scalars $\bar{\delta}_\sigma$ and $\underline{\delta}_\sigma$ that satisfy (20) and (19), such that

$$\bar{z}_\sigma(\chi) - \underline{z}_\sigma(\chi) \leq \bar{\delta}_\sigma + \underline{\delta}_\sigma. \tag{26}$$

Integrating both sides of this relation over $\sigma \in [0, T - t_c(\chi) - t_d(\chi)]$ yields

$$A(\chi) = \int_0^{T-t_c(\chi)-t_d(\chi)} \bar{z}_\sigma(\chi) - \underline{z}_\sigma(\chi) d\sigma \leq (C_{\max} - C)T \tag{27}$$

Since $C \leq C_{\max}$, the condition in (27) is satisfied for $\chi \in [-C/2, C/2]$ if and only if it is also satisfied on $\{\chi : A(\chi) \geq 0\}$. On this set, $A(\chi)$ equals the area of the parallelogram in Figure 8 with base and height

$$B(\chi) = (1-\frac{P_0}{\bar{P}})T - t_d(\chi)$$
$$= \left(1-\frac{P_0}{\bar{P}}\right)T\left(1-\frac{C/2+\chi}{\underline{w}C_{\max}}\right)$$
$$H(\chi) = E - \bar{P}t_c(\chi) = E\left(1-\frac{C/2-\chi}{\bar{w}C_{\max}}\right).$$

Hence, (27) holds on $\{\chi \in [-C/2, C/2] : A(\chi) > 0\}$ if and only if

$$a(\chi') = b(\chi')h(\chi') \leq 1-c, \text{ for all } \chi' \in \mathbb{I}, \tag{28}$$

where $b(\chi) = (1-\frac{c/2+\chi'}{\underline{w}})$, $h(\chi') = (1-\frac{c/2-\chi'}{\bar{w}})$ and

$$\mathbb{I} = [-c/2, c/2] \cap \{\chi' : b(\chi), h(\chi') \geq 0\}$$
$$= [-c/2, c/2] \cap [c/2 - \underline{w}, -c/2 + \bar{w}].$$

The extremum

$$\max_{\chi' \in \mathbb{R}} a(\chi') = \frac{(\bar{w} + \underline{w} - c)^2}{4\bar{w}\underline{w}}$$

is attained at $\chi^* = (\underline{w}-\bar{w})/2$. The set $\mathbb{I}$ is non-empty if and only if $\bar{w}+\underline{w} \geq c$. In this case, it is straightforward to verify that $\chi^* \in \mathbb{I}$ if and only if $|\bar{w}-\underline{w}| \leq c$. We conclude that if $\mathbb{B}(\phi) \subset \mathbb{W}$ then (12) holds on

$$\{c, \bar{w}, \underline{w} \in [0,1]: \bar{w}+\underline{w} \geq c, |\bar{w}-\underline{w}| \leq c\}.$$

This set contains $\mathbb{S}$.

### E. Proof of Proposition 3 (and Proposition 4)

It follows from Lemma 3 that $x_\sigma \geq \bar{z}_\sigma(\chi)$. Combining this with (7) yields $x_\sigma \geq \max(\bar{z}_\sigma(\chi), \underline{x}_\sigma)$. Since, by assumption, $\chi = x_{\text{avg}}$, we have

$$\chi = \frac{1}{T}\int_0^T x_\sigma - x_\sigma^{\text{nom}} d\sigma \geq \frac{1}{T}\int_0^{T-t_c} \bar{z}_\sigma - \underline{x}_\sigma d\sigma$$
$$-\frac{1}{T}\int_0^T x_\sigma^{\text{nom}} - \underline{x}_\sigma d\sigma = (E-\bar{P}t_c)(1-\frac{P_0}{\bar{P}}) - \frac{C_{\max}}{2} = \chi \tag{29}$$

so $x_\sigma = \max(z_\sigma, \bar{x}_\sigma)$, which satisfies the assertion. In the second to last equality in (29) we have used both Lemma 2 and $\int(\bar{z}_\sigma - \underline{x}_\sigma) = (E-\bar{P}t_c)(1-\frac{P_0}{\bar{P}})T$, which is straightforward to derive using Figure 8. The proof of Proposition 4 is analogous.

### F. Proof of Theorem 2

To prove that (18) is sufficient for $\mathbb{B}(\phi) \subset \mathbb{W}(\mu^\eta)$, suppose that $w \in \mathbb{B}(\phi)$ and set $p_\tau = \mu_\tau^\eta(w)$. We must show that (1) and

$$\frac{1}{T}\int_\mathbb{R} \mu_\tau(w) d\tau = \frac{1}{T}\int_\mathbb{R} p_\tau^{\text{nom}} d\tau + w \tag{30}$$

are satisfied. Let $x_\sigma^\eta(\chi) = [y_\sigma(\chi)]_{\underline{x}_\sigma}^{\bar{x}_\sigma}$, where

$$y_\sigma(\chi) = \eta(E - \bar{P}T + \bar{P}\sigma) + \bar{P}s(\chi),$$

and $s(\chi)$ is such that

$$\frac{1}{T}\int_0^T (x_\sigma^\eta(\chi) - x_\sigma^{\text{nom}}) d\sigma = \chi.$$

Figure 4 shows $x^\eta(0)$ for some different $\eta$. We claim that under the assumptions $x(t) = x^\eta(\chi(t))$ for all $t \geq 0$. Suppose that $x(t) = x^\eta(\chi(t))$ for some $t$ and define

$$\mathbb{I}(\chi) = \{\sigma : x_\sigma = y_\sigma(\chi)\}.$$

It is straightforward to verify that if $\sigma \in \mathbb{I}(\chi)$, then $s_\sigma^\eta = s(\chi)$. Moreover, if $x_\sigma$ is smaller (larger) than $y_\sigma(\chi)$, then $s_\sigma^\eta$ is smaller (lager) than $s(\chi)$. Hence, $\mu^\eta$ will attempt to absorb $w$ by first increasing or decreasing $x_\sigma$, $\sigma \in \mathbb{I}(\chi)$. For these loads, we have $\partial x_\sigma(t)/\partial \sigma = \eta \overline{P}$. Using (6), we have

$$\dot{s}^\eta_\sigma(t) = \dot{x}_\sigma(t)/\overline{P} = -\eta + p_{t-\sigma}(t)/\overline{P} \text{ for all } \sigma \in \mathbb{I}(\chi(t)).$$

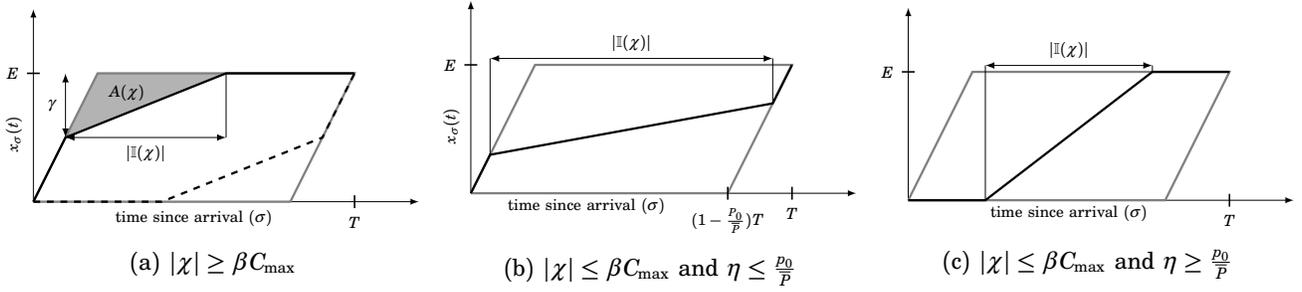

Fig. 9: Illustration of $x^\eta$. In Figure 9a the solid and dashed curves illustrate the case for $\chi \geq \beta C_{\max}$ and $\chi \leq -\beta C_{\max}$, respectively, and due to symmetry, $\mathbb{I}(\chi)$ is independent of the sign of $\chi$.

It follows by the definition of $\mu^\eta$, that $p_{t-\sigma}(t) = p_{t-\sigma'}(t)$ for all $\sigma, \sigma' \in \mathbb{I}(\chi)$. Also, $\dot{x}_\sigma(t) = 0$ implies $p_{t-\sigma}(t) = \eta \overline{P}$. It is straightforward to verify that, if $\chi(t) \in [-C/2, C/2]$, then

$$\frac{1}{T} \int_{\mathbb{I}} \dot{x}_\sigma(t) d\sigma = w(t) \iff w(t) \in \underbrace{\frac{|\mathbb{I}(\chi)|\overline{P}}{T}[-\eta, (1-\eta)]}_{\mathbb{A}(\chi)}$$

where $|\mathbb{I}(\chi)|$ is the length of the interval $\mathbb{I}(\chi)$. To find an expression for $|\mathbb{I}(\chi)|$, set $\alpha = \frac{(1-\eta)P_0}{\eta(\overline{P}-P_0)}$ and $\beta = \frac{\max\{1-\alpha, 1-\alpha^{-1}\}}{2}$.

For $|\chi| \geq \beta C_{\max}$, $x^\eta$ is shown in Figure 9a. Using

$$\frac{\gamma}{2}\left(|\mathbb{I}(\chi)| - \frac{\gamma}{\overline{P}}\right) = A(\chi) = \left(\frac{C_{\max}}{2} - \chi\right)T$$
$$\eta \overline{P} |\mathbb{I}(\chi)| = \gamma$$

we obtain $\frac{|\mathbb{I}(\chi)|\overline{P}}{T} = \sqrt{\frac{\kappa(\chi)\overline{W}\underline{W}}{\eta(1-\eta)}}$, where $\kappa(\chi) = \frac{1-2\chi/C_{\max}}{\overline{w}\underline{w}}$. From (18) it follows that $\kappa(\chi) \geq 1$ for all $\chi \in [-C/2, C/2]$. Using

$$\eta/(1-\eta) = \underline{W}/\overline{W} \tag{31}$$

yields $\mathbb{A}(\chi) \supset [-\underline{W}, \overline{W}]$.

Now consider $|\chi| \leq \beta C_{\max}$. If $\eta \leq P_0/\overline{P}$, $x^\eta$ is illustrated in Figure 9b. It satisfies

$$\eta \overline{P} |\mathbb{I}| = (|\mathbb{I}| - (1 - P_0/\overline{P})T)\overline{P} \implies \frac{|\mathbb{I}|\overline{P}}{T} = \frac{\overline{W}_{\max}}{(1-\eta)}.$$

Using this together with (31) yields $\mathbb{A}(\chi) \supset [-\underline{W}, \overline{W}_{\max}]$. Similarly, if $\eta \leq P_0/\overline{P}$ (Figure 9c), we have $\frac{|\mathbb{I}(\chi)|\overline{P}}{T} = \frac{\underline{W}_{\max}}{\eta}$, from which it follows that $\mathbb{A}(\chi) \supset [-\underline{W}_{\max}, \overline{W}]$.

We conclude that $\mathbb{A}(\chi) \supset [-\underline{W}, \overline{W}]$ for all $\chi \in [-C/2, C/2]$. This implies that $\dot{x}_\sigma(t) = 0$ for $\sigma \notin \mathbb{I}(\chi)$, and by the definition of $s(\chi)$ that $\dot{x}_\sigma(t) = \frac{dy_\sigma(\chi(t))}{dt}$ for $\sigma \in \mathbb{I}(\chi)$. The claim follows by noting that $x^\eta(0)$ is an equilibrium state under $\mu^\eta$, so $x(0) = x^\eta(0)$.

From $x = x^\eta(\chi(t))$ it follows immediately that (1a), (1b) and (30) are satisfied for all $w \in \mathbb{B}(\phi)$. From $\mathbb{A}(\chi) \supset [-\underline{W}, \overline{W}]$ it also follows that (1c) is satisfied.

For the necessity assertion, suppose that $\overline{w} = 1$, but that $\underline{w} + c > 1$, which violates (18). In this case, the assumptions of Theorem 1 are satisfied, and (12) reduces to $(\underline{w}+c-1)^2 \leq 0$. This implies $\underline{w}+c = 1$, which is a contradiction. By symmetry, the same arguments can be used to show that (18) is necessary when $\underline{w} = 1$.

Next, suppose that $c = 1$, $\overline{w} > 0$ and $\chi(t) = \overline{C}_{\max}/2$ for some $t$. From the definition of $\overline{z}$ in (22) it follows that $x(t) = \overline{x}(t) = \overline{z}(t)$. Hence, $\dot{x}_\sigma \geq -\partial x_\sigma/\partial \sigma = 0$ for $\sigma > E/\overline{P}$, and Lemma 3 implies $\dot{x}_\sigma \geq \dot{z}_\sigma = 0$ for $\sigma \in [0, E/\overline{P}]$. Hence, the stored energy must satisfy

$$\dot{x}_{\text{avg}}(t) = \frac{1}{T} \int_0^T \dot{x}_\sigma(t) d\sigma \geq 0 \implies \underline{w} = 0.$$

The steps to show that $\overline{w} = 0$ if $c = 1$ and $\underline{w} > 0$ are analogous.